\documentclass[12pt]{article}
\usepackage{epsfig, amsmath, amssymb}
\usepackage{graphicx,psfrag} 
\newcommand{\be}{\begin{equation}}
\newcommand{\ee}{\end{equation}}
\newcommand{\bea}{\begin{eqnarray}}
\newcommand{\eea}{\end{eqnarray}}
\newcommand{\bA}{\begin{array}}
\newcommand{\eA}{\end{array}}
\newcommand{\bc}{\begin{center}}
\newcommand{\ec}{\end{center}}
\newcommand{\al}{\alpha}

\newcommand{\ra}{\rightarrow}
\newcommand{\del}{\partial}

\newcommand{\ie}{{\it i.e.}}
\newcommand{\eg}{{\it e.g.}}

\newcommand{\Nf}{${\cal N}{=}4$}

\def\BZ{{\mathbb Z}}

\begin{document}


\begin{titlepage}
\vspace{30mm}

\bc

\hfill 
\\         [22mm]

{\huge Lifshitz-like systems \\ [2mm] 
and $AdS$ null deformations}
\vspace{16mm}

{\large K.~Narayan} \\
\vspace{3mm}
{\small \it Chennai Mathematical Institute, \\}
{\small \it SIPCOT IT Park, Padur PO, Siruseri 603103, India.\\}

\ec
\medskip
\vspace{40mm}

\begin{abstract}
Following arXiv:1005.3291 [hep-th], we discuss certain lightlike 
deformations of $AdS_5\times X^5$ in Type IIB string theory sourced 
by a lightlike dilaton $\Phi(x^+)$ dual to the \Nf\ super Yang-Mills 
theory with a lightlike varying gauge coupling. We argue that in the 
case where the $x^+$-direction is noncompact, these solutions describe 
anisotropic 3+1-dim Lifshitz-like systems with a potential in the 
$x^+$-direction generated by the lightlike dilaton. We then describe 
solutions of this sort with a linear dilaton. This enables a detailed 
calculation of 2-point correlation functions of operators dual to 
bulk scalars and helps illustrate the spatial structure of these 
theories. Following this, we discuss a nongeometric string 
construction involving a compactification along the $x^+$-direction
of this linear dilaton system. We also point out similar IIB axionic 
solutions. Similar bulk arguments for $x^+$-noncompact can be carried 
out for deformations of $AdS_4\times X^7$ in M-theory.
\end{abstract}

\end{titlepage}

{\tiny 
\begin{tableofcontents}
\end{tableofcontents}
}

\vspace{2mm}

\section{Introduction}

It is interesting to explore generalizations of holographic duality
to physical systems with nonrelativistic symmetries, with a view 
towards possible interfaces with condensed matter systems (see \eg\ 
\cite{Hartnoll:2009sz,Herzog:2009xv,McGreevy:2009xe} for reviews).
In this paper, we discuss Lifshitz-like fixed points from a holographic
point of view.

Lifshitz points arise in various condensed matter systems, \eg\
magnetic systems with antiferromagnetic interactions, dimer models,
liquid crystals and so on (see \eg\ \cite{chaikinluben,cardy}
for lucid descriptions). They exhibit the anisotropic 
scaling\  $t\ra \lambda^z t,\ x_i \ra \lambda x_i$, with $z$ the 
dynamical exponent. A Landau-Ginzburg description for such theories 
with $z=2$ has the effective action\ 
$S = \int dt d^2x\ ((\del_t\varphi)^2 - \kappa (\nabla^2\varphi)^2)$.
Aspects of correlation functions in these theories have been 
discussed in \eg\ \cite{ardfendfradkin,ashvinSenthilLifsh}. 

Holographic dual gravitational models of such theories were described
in \cite{Kachru:2008yh}: these spacetimes\ $ds^2=-{dt^2\over r^{2z}} +
{dx_i^2 + dr^2 \over r^2},\ i=1,2,$\ arise as solutions to 4-dim
Einstein gravity with a cosmological constant coupled to a massive
abelian gauge field.  The symmetries exhibited by this spacetime are
time translations, spatial translations/rotations, as well as the
anisotropic scaling above: these are smaller than the Galilean
symmetries explored holographically in
\cite{Son:2008ye}-\cite{Donos:2009zf}.  Previous attempts at string or
supergravity constructions of such spacetimes $Lif_4$ include \eg\
\cite{Taylor:2008tg,Li:2009pf,
  Blaback:2010pp,koroteev,Azeyanagi:2009pr,Hartnoll:2009ns}.

In \cite{Balasubramanian:2010uk}, certain lightlike deformations of
$AdS_5\times X^5$ sourced by a lightlike dilaton in Type IIB string
theory (as well as those of $AdS_4\times X^7$ in M-theory) were argued
to give rise, upon dimensional reduction, to $z=2$ Lifshitz spacetimes
in 3+1- and 2+1-dimensions.  For the $AdS_5$ case, these are dual to a
DLCQ of the \Nf\ super Yang-Mills theory with a gauge coupling that
varies along the compact direction.  Some of the supporting evidence
includes symmetry arguments from both the bulk and dual field theory
points of view, as well as matching of certain equal-time 2-point
correlation functions with those found in \cite{Kachru:2008yh}. These
constructions were found to nicely generalize \cite{Donos:2010tu} to a
large family of similar $z=2$ solutions with various other fields
incorporated.  Lifshitz-like solutions with more general values of the
dynamical exponent $z$ have been constructed in \cite{Gregory:2010gx}
(see also related work in
\cite{Donos:2010ax,Singh:2010zs,Costa:2010cn,Nishioka:2010ha,Cassani:2011sv,
  Halmagyi:2011xh}).

In this paper, we continue to explore the apparently simpler systems
in \cite{Balasubramanian:2010uk}, but now without any
compactification.  This is a lightlike deformation of $AdS_5\times
X^5$ sourced by a null dilaton, dual to the \Nf\ SYM with a gauge
coupling varying along one lightlike direction. This system turns out
to be interesting and exhibits spatially anisotropic $3+1$-dim
Lifshitz-like symmetries with dynamical exponents $z=2$ in the
$x_i,x^-$-directions and $z=\infty$ in the $x^{\pm}$-directions. The
metric and dilaton respect the scaling symmetry but break
$x^+$-translation invariance. In addition, the lightlike dilaton
configuration gives rise to a potential in the $x^+$-direction.

For the particular case of a dilaton that is linear in the lightlike
$x^+$-direction, the bulk Einstein metric becomes independent of $x^+$
and the spacetime simplifies. In particular, this enables a detailed
calculation of the 2-point correlation function for operators dual to
bulk scalar modes. The resulting structure obtained from this
$AdS/CFT$ calculation bears some similarity to that found in the
effective gravity Lifshitz hologram of \cite{Kachru:2008yh}. However
there is further structure in this case due to the linear dilaton
configuration, which is reminiscent of Liouville-like walls in
theories in $c=1$ string theory. The linear dilaton $x^+$-potential
has some reflection in the 2-point momentum space correlation function
for massless scalars (dual to dimension-4 operators) which contains
some structure resembling a mass-gap in the $x^+$-direction. This
suggests that solutions of this form will in general exhibit features
reflecting the spatial $x^+$-potential generated by the dilaton.

Much of this bulk discussion for $x^+$-noncompact can be carried out
for similar lightlike deformations of $AdS_4\times X^7$ in M-theory,
giving insight into $2+1$-dim Lifshitz-like field theories possibly
dual to lightlike deformations of Chern-Simons theories arising on
M2-branes holographically \cite{M2}.

We then discuss a possible dimensional reduction of a linear dilaton
system involving a nongeometric string construction using S-duality of
the IIB theory. We also point out very similar Lifshitz-like solutions
sourced by the axion in IIB string theory. We finally close with a
discussion including comments on some specific solutions.

\section{Reviewing $z=2$ Lifshitz spacetimes from $AdS$ 
null deformations with null dilaton}

In \cite{Balasubramanian:2010uk}, we studied null deformations of
$AdS\times X$ sourced by a lightlike dilaton in 10- or 11-dim
supergravity or string (or M-) theory, and argued that upon
dimensional reduction (DLCQ) they represent gravity duals of $z=2$
Lifshitz fixed points in 2+1- or 1+1-dimensions. The 10- or 11-dim 
bulk system with $x^+$-compact is
\be\label{metLif5}
ds^2 = {1\over w^2} [-2dx^+dx^- + dx_i^2 + \gamma w^2 (\Phi')^2 (dx^+)^2]
+ {dw^2\over w^2} + d\Omega_S^2\ ,\qquad \Phi=\Phi(x^+)\ ,
\ee
with a corresponding 5- or 4-form field strength. The constant $\gamma$ 
is $\gamma={1\over 4}$ for $AdS_5$ and $\gamma={1\over 2}$ for $AdS_4$
(the $d\Omega_S^2$ is the metric for $S^5$ or $X^7$ respectively, with 
$X^7$ being some Sasaki-Einstein 7-manifold).\\
It is natural to interpret $x^-$ as the time variable since a
constant-$x^-$ surface is spacelike (since $g^{--}<0$), while a
constant-$x^+$ surface is null.
The spacetime (\ref{metLif5}) exhibits the following symmetries: 
translations and rotations in $x_i$, translations in $x^-\equiv t$ 
(time), and a $z=2$ scaling\ 
$x^-\ra \lambda^2 x^- , x_i\ra \lambda x_i , w\ra \lambda w$\ ($x^+$ 
being compact does not scale). Possible Galilean boosts\ 
$x_i\ra x_i-v_ix^- ,\ x^+\ra x^+-{1\over 2} (2v_ix_i-v_i^2x^-)$,\ are 
broken by the $g_{++}\sim (\Phi')^2$ term.  (If $g_{++}=0$, this is 
essentially $AdS$ in lightcone coordinates and the system upon 
DLCQ has a Schrodinger symmetry, as discussed in \eg\ 
\cite{Goldberger:2008vg,Barbon:2008bg,Maldacena:2008wh}).

In the $AdS_5$-deformed case, the gauge theory dual to these systems
can be identified to be the DLCQ of the \Nf\ super Yang-Mills theory
with a gauge coupling varying along the $x^+$-direction as\
$g_{YM}^2=e^{\Phi(x^+)}$. The boundary metric\ $\lim_{w\ra 0} ds_4^2$\
is flat, so the gauge theory lives on flat spacetime.  From the point
of view of the dual gauge theory, the symmetry structure is
intuitively clear: noting that the DLCQ of a relativistic field theory
gives a nonrelativistic (Galilean) system, we see that the gauge
coupling varying along the $x^+$-direction then breaks the $x^+$-shift
symmetry reducing the Galilean symmetry down to a Lifshitz one.

It can be checked directly that these spacetimes (\ref{metLif5}) along with 
the scalar $\Phi$ and appropriate 5-form (or 4-form) field strength 
are solutions to the 10-dim (or 11-dim) supergravity equations. For 
instance, there is no $S^5$ or $X^7$ dependence and the resulting 5- 
or 4-dim system, with an effective cosmological constant from the flux, 
solves the equations\ $R_{MN}=-d g_{MN} + {1\over 2} \del_M\Phi\del_N\Phi$, 
with $d=4,3,$ for $AdS_{d+1}$, being the 5- or 4-dim effective 
cosmological constant. Finally the lightlike nature ensures that the 
scalar equation of motion is automatically satisfied.\\
However it is worth mentioning that the coordinate transformation\ 
$w=r e^{-f/2} , \ x^-=y^--{w^2 f'\over 4} $,\ recasts these spacetimes 
(\ref{metLif5}) into the form (we set the AdS radius $R=1$)
\be\label{metconf5}
ds^2={1\over r^2} [e^{f(x^+)} (-2dx^+dy^- + dx_i^2) + dr^2] + d\Omega_5^2, 
\qquad \Phi=\Phi(x^+) \ ,
\ee
with the 4-dim part being conformal to flat space, the boundary 
metric becoming\ $e^f \eta_{\mu\nu}$: indeed this is where the 
$AdS_5$-deformed systems were originally found 
\cite{Das:2006dz,Das:2006pw,Awad:2007fj,Awad:2008jf} (see also 
\cite{Chu:2006pa,Lin:2006ie})\footnote{After this paper appeared on 
the arXiv, we were informed of earlier solutions representing null 
deformations of $AdS_3$ in the context of WZW models \cite{orlando}.}. 
These are Penrose-Brown-Henneaux (PBH)
transformations, a subset of bulk diffeomorphisms leaving the metric
invariant (in Fefferman-Graham form), and acting as a Weyl
transformation on the boundary. We will refer to the coordinate system
in (\ref{metconf5}) as conformal coordinates in what follows. The only
nonzero Ricci component is $R_{++}$, giving\ 
$R_{++} = {1\over 2} (\del_+\Phi)^2$, \ie\ 
$R_{++} = {1\over 2} (f')^2 - f'' =  2\gamma (\Phi')^2$ ,
with $\Phi'\equiv {d\Phi\over dx^+}\ ,\ f'={df\over dx^+}$\ .\\
The $AdS_5$-deformed solutions were shown to preserve half (lightcone) 
supersymmetry in the form (\ref{metconf5}) in \cite{Das:2006dz}. 
Supersymmetry was also shown for various solutions in the more 
general family in \cite{Donos:2010tu} which are generalizations 
of the metric form (\ref{metLif5}).

An argument for the dimensional reduction of the spacetime
(\ref{metLif5}) along the compact $x^+$-direction was given in
\cite{Balasubramanian:2010uk}.  This suggests that the system has the
right structure, although a clear Wilsonian separation-of-scales
argument allowing for a standard Kaluza-Klein reduction of this metric
is difficult due to the $x^+$-dependence of $g_{++}$ in
(\ref{metLif5}) for generic $\Phi(x^+)$. 

Further checks involve the equal-time 2-point correlation function of
operators dual to bulk scalar modes: the bulk calculation in this
spacetime agrees with the spatial power-law behaviour noted by
\cite{Kachru:2008yh}, as we review briefly now.
In the conformal coordinates (\ref{metconf5}), the holographic 2-point
correlation function for operators dual to scalars can be found in
closed form \cite{Das:2006pw}: this was used in
\cite{Balasubramanian:2010uk} to show agreement of the equal-time
expression\ ${1\over [(\Delta x_i)^2]^\Delta}$\ with that in
\cite{Kachru:2008yh}.
The conformally flat boundary metric is $e^{f(x^+)} \eta_{\mu\nu}$ and 
the boundary coupling of the bulk mode $\varphi$ is\ 
$\int d^3x dx^+ e^{2f(x^+)} {\cal O} \varphi$.
Then the holographic boundary action is
\bea\label{conf2ptAction}
S &=& C \int d^4x d^4x'\ e^{3f(x^+)/2} e^{3f({x^+}')/2}\
\varphi(x^+,{\vec x}) \varphi({x^+}',{\vec x'})\
\biggl(\frac{\Delta\lambda}{\Delta x^+}\biggr)^{1-\Delta}\ \frac{1}{
[(\Delta{\vec x})^2]^{\Delta}}\ , \nonumber\\ [2mm]
&& \ (\Delta {\vec x})^2 = -2(\Delta x^+)(\Delta x^-) 
+ \sum_{i=1,2} (\Delta x^i)^2\ , \qquad  \Delta=2+\nu=2+\sqrt{4+m^2}\ .
\eea
where $C$ is a constant,\ $\lambda=\int e^{f(x^+)} dx^+$ is 
the affine parameter along null geodesics stretched solely along $x^+$, 
and $(\Delta {\vec x})^2$ is the 4-dimensional distance element.
As it stands, this is a 4-dim field theory boundary action: to obtain 
an action for an effective 3-dim boundary field theory, we need to 
dimensionally reduce over the $x^+$-direction. In the compactified 
limit\ $\Delta x^+\ll \Delta x^-, \Delta x_i$,\ it is consistent to 
approximate\ ${\Delta\lambda\over\Delta x^+}\sim {d\lambda\over dx^+}=e^f$,\ 
and $e^{f(x^+)}\sim 1$, essentially smearing the $x^+$ dependence 
relative to the uncompactified dimensions. In this approximation, 
we can read off the equal-time 2-point function as\ 
$\langle {\cal O}(x_i) {\cal O}(x'_i)\rangle\ \sim\ 
{1\over [\sum_i (\Delta x^i)^2]^{\Delta}}$ . Likewise for two points 
at essentially the same spatial location (small $\Delta x_i$), we 
have\ $\langle {\cal O}(t){\cal O}(t')\rangle\sim\ 
{1\over (\Delta x^-)^{\Delta}}$ ,\ where we have suppressed the 
$x^+$-integrals. The additional $x^+$-dependences distinguish 
this from a Galilean theory arising from a DLCQ. This power law 
spatial and temporal falloff behaviour and their associated scaling
again vindicates the $z=2$ Lifshitz scaling: similar power law 
behaviour was exhibited for some simple operators in the free
Lifshitz field theory in \cite{ardfendfradkin}.\\
The corresponding calculation in the metric (\ref{metLif5}) appears 
relatively difficult to do since the $g_{++}$ piece is $x^+$-dependent 
for general $\Phi(x^+)$ and ruins a simple separation of variables 
approach to solve the scalar wave equation.

The 11-dim $z=2$ Lifshitz-like solutions in (\ref{metLif5}) 
involve a dimensional reduction of null deformations of $AdS_4\times
X^7$, with $X^7$ being some Sasaki-Einstein 7-manifold. In this case,
the scalar does not have any natural interpretation in the 11-dim
theory directly: it arises instead from the 4-form flux after
compactification on $X^7$.  We expect that these are dual to the DLCQ
of appropriate lightlike deformations of Chern-Simons theories arising
on M2-branes holographically \cite{M2} (and various generalizations).

We mention in passing that there also exist time dependent
deformations of $AdS_5$ \cite{Das:2006dz,Awad:2007fj,Awad:2008jf} and
$AdS_4$: in particular the asymmetric Kasner-like solutions exhibit
interesting (anisotropic) Lifshitz scaling symmetries. These solutions
are qualitatively different from the null ones above.

Finally \cite{Balasubramanian:2010uk} also discussed a solution of
5-dimensional gravity with negative cosmological constant and a
massless complex scalar, that are similar to the null solutions
(\ref{metLif5}) above: these upon dimensional reduction give rise to
$2+1$-dim Lifshitz spacetimes. This 5-dim solution can be uplifted to
11-dimensional supergravity.

\section{$x^+$-noncompact and anisotropic Lifshitz systems}

We want to now study the IIB null-deformed system (\ref{metLif5}) 
\be\label{metLif55}
ds^2 = {1\over w^2} [-2dx^+dx^- + dx_i^2 + {1\over 4} w^2 (\Phi')^2 (dx^+)^2]
+ {dw^2\over w^2}\ ,\qquad \Phi=\Phi(x^+)\ , 
\ee
but with $x^+$ treated as a noncompact direction. 
Firstly it is worth noting that the functional dependence of the 
dilaton is really\ $\Phi=\Phi(Qx^+)$, the constant $Q$ being 
a parameter of mass dimension one. Lightlike boosts\ 
$x^+\ra\lambda x^+,\ x^-\ra\lambda^{-1} x^-$,\ were symmetries in 
the original system with $Q=0$:\ these are broken in the present
case. These boosts rescale $Q$ as\ $Q\ra {Q\over\lambda}$ so that
theories with different values of $Q$ are related by these boosts.

In this case, the symmetries include time $x^-$-translations, spatial
$x_i$-translations/rotations: translations in the $x^+$-direction are 
broken by the nontrivial $x^+$-dependence. In addition, the metric 
(\ref{metLif55}) exhibits the scaling
\be\label{z=2infty}
w\ra \lambda w ,\quad x_i\ra \lambda x_i ,\quad 
x^-\ra \lambda^2 x^- , \quad x^+\ra \lambda^0 x^+ .
\ee
The system of course contains the $z=2$ Lifshitz scaling symmetry\
$x_i\ra \lambda x_i ,\ \ x^-\ra \lambda^2 x^-$\ \ in the $2+1$-dimensions 
$x^-,x_i$\ (induced by the associated scaling of $w$).
However in addition note that since $x^+$ does not scale, we effectively 
have $z=\infty$ Lifshitz scaling in the $x^+,x^-$-directions (reading 
off the dynamical exponent as the ratio of the scaling of time ($x^-$) 
to the spatial one ($x^+$)). Thus it appears best to interpret this 
system as a spatially anisotropic Lifshitz 
system, with $z=2$ scaling for the $2+1$-dimensional $x_i,x^-$-plane 
and $z=\infty$ scaling for the $x^+,x^-$-directions. This is reminiscent 
of the anisotropic scaling\footnote{We recall that radial Kasner-like 
solutions of the form\ 
$ds^2 = {1\over r^2} [dr^2 - r^{2p_0} dt^2 + \sum_i r^{2p_i} (dx^i)^2 ]$ 
exist, sourced by several massive vector fields \cite{Taylor:2008tg}.} 
observed in the D3-D7 construction of \cite{Azeyanagi:2009pr} and 
the scalings in the dilatonic black brane solutions in 
\cite{Goldstein:2009cv}: however in this case, the scaling 
(\ref{z=2infty}) is an actual symmetry of (\ref{metLif55}), 
respected by the dilaton $\Phi(x^+)$ as well. This sort of 
anisotropic scaling would seem to also hold for some of the more 
general solutions in \cite{Donos:2010tu}. The dilaton however 
breaks $x^+$-translation invariance, and in fact gives rise to a 
spatial potential in the $x^+$-direction: this gives rise to 
additional structure in observables in this system which reflect 
this effective $x^+$-potential. In the next subsection, we will 
analyse the case of a linear dilaton potential in some detail, 
illustrating some of this structure.

The system (\ref{metLif55}) also exhibits the symmetry\ $x_i\ra
x_i-v_ix^+$ with $x^+$ unchanged and a corresponding shift in $x^-$:\
this is broken if $x^+$ is compact. This symmetry however is not a
Galilean boost since $x^+$ cannot be interpreted as time:\ $x^-$ is
the natural time coordinate here, consistent with constant-$x^-$
surfaces being spacelike ($g^{--}\sim\ -w^4(\Phi')^2<0$). This sign of
$g^{--}$ appears crucial for this interpretation and the $z=\infty$
scaling we have mentioned above. Some of the more general solutions in
\cite{Donos:2010tu} have $g_{++}<0$, and are more akin to $z=0$
Schrodinger systems\ $ds^2=-dt^2+{dx_i^2+dt d\xi+dr^2\over r^2}$ .
Indeed transforming $t\ra ix^+ ,\ \xi\ra ix^-$ recast this solution
into the form in (\ref{metLif55}) with $\Phi'=const$. Thus its
symmetries formally exist for (\ref{metLif55}) too.
In the present case, it appears best to interpret (\ref{metLif55}) as
an anisotropic Lifshitz-like system with a spatial $x^+$-potential
stemming from the dilaton. This is corroborated by our discussion on
linear dilatonic systems in the next subsection, where we also
calculate some correlation functions which in part exhibit structure
similar to those in \cite{Kachru:2008yh}. There we will also make
further comments on this point.

As we have mentioned, these can be recast in conformal coordinates 
(\ref{metconf5}), with conformally flat boundary metric 
$e^{f(x^+)}\eta_{\mu\nu}$. In these variables, the holographic 2-pt fn 
for operators ${\cal O}(x)$ with boundary coupling\ 
$\int d^4x e^{2f(x^+)} {\cal O}(x) \varphi(x)$\ to the bulk mode 
$\varphi(x)$ can be read off from (\ref{conf2ptAction}) as \cite{Das:2006pw}
\be\label{2ptconf}
\langle O(x) O(x')\rangle = e^{-f(x^+)/2} e^{-f({x'}^+)/2}
\left(\frac{\Delta\lambda}{\Delta x^+}\right)^{1-\Delta} 
\frac{1}{[(\Delta{\vec x})^2]^{\Delta}} \ ,
\ee
with\ $\Delta=2+\sqrt{4+m^2},\ \ \lambda=\int e^{f(x^+)} dx^+$.
For two points with $\Delta x^+\ll \Delta x^-, \Delta x_i$, \ie\ that 
are essentially on a constant-$x^+$ slice, we can approximate 
${\Delta\lambda\over\Delta x^+}\sim {d\lambda\over dx^+}=e^f$, and 
$(\Delta{\vec x})^2 \sim\ (\Delta x_i)^2$. This gives\
$\langle O(x) O(x')\rangle
 \sim\ \frac{e^{-f(x^+)\Delta}}{[\sum_i (\Delta x_i)^2]^{\Delta}}$ .
The factor $e^{-f\Delta}$ here is a reflection of the fact that the 
conformally dressed operators $e^{f(x^+)\Delta/2} {\cal O}(x)$ in this 
conformally flat background $e^f\eta_{\mu\nu}$ behave like undressed 
operators in the flat space background, possessing a flat space 
2-point function.
We also see that the equal time correlator ($\Delta x^-=0$) is\
$({\Delta\lambda\over\Delta x^+})^{1-\Delta}\ 
{e^{-f(x^+)\Delta/2} e^{-f({x'}^+)\Delta/2}\over [\sum_i (\Delta x_i)^2]^{\Delta}}$ ,
exhibiting spatial power-law behaviour in the $x_i$, similar to 
the equal time 2-pt correlator in \cite{Kachru:2008yh}, but also 
possessing additional $x^+$-dependence.
Note that this calculation has been done at the boundary $r=\epsilon$:\
recalling that the radial coordinates are related as\ $w=r e^{-f(x^+)/2}$,
this boundary differs from the corresponding boundary $w=\epsilon$ in 
the metric (\ref{metLif55}), although they are in the same conformal 
class.

In the next subsection, we discuss the case of a linear dilaton, in 
which case the holographic correlator can be calculated in the 
metric (\ref{metLif55}), giving some detailed insight into the 
structure of this system.

\subsection{Linear-dilaton-like deformations}

As we have seen, the AdS null-deformed solutions (\ref{metLif55}) have 
$g_{++}={(\Phi')^2\over 4}$ .\ For $\Phi'=const$, we see that the Einstein 
metric is independent of $x^+$. This is the case of a dilaton that is 
linear. 

Consider $\Phi'=const$: this gives $\Phi=\Phi_0+2Qx^+$, which is a
linear dilaton profile, the constant $Q$ being a parameter (we have
chosen the constant $2Q$ for convenience).
Then the bulk spacetime and dilaton are
\be\label{metLinDil}
ds^2 = {1\over w^2} [-2dx^+dx^- + dx_i^2 + w^2 Q^2 (dx^+)^2] 
+ {dw^2\over w^2}\ ,\qquad \Phi=\Phi_0+2Qx^+\ .
\ee
The symmetries in this case, besides those mentioned above 
(\ref{z=2infty}), also include translations in $x^+$ in the metric 
(\ref{metLinDil}): however there is a spatial $x^+$-potential 
stemming from the linear dilaton.

The action for a massless scalar\ 
$S={1\over G_5} \int d^5x\ \sqrt{-g}\ g^{\mu\nu}\del_\mu\varphi\del_\nu\varphi$\
on restricting to modes propagating on a constant-$x^+$ surface, \ie\ 
with no $x^+$-dependence ($\del_+\varphi=0$), 
\bea
S &=& {1\over G_5} 
\int {d^4x dx^+\over w^5} \left[ -{w^4 (\Phi')^2\over 4} (\del_-\varphi)^2 
- 2w^2 (\del_-\varphi)(\del_+\varphi) + w^2 (\del_i\varphi)^2 
+ w^2 (\del_w\varphi)^2 \right]\nonumber\\
&& = {1\over G_5} \int dx_+ {d^4x\over w^5} 
\left[ - Q^2 w^4 (\del_-\varphi)^2 
+ w^2 (\del_i\varphi)^2 + w^2 (\del_w\varphi)^2 \right]\ .
\eea
Such scalar modes see an effective $z=2$ Lifshitz geometry in the 3+1-dim 
($x^-,x_i,w$) part of the bulk. Modes propagating only in the 2+1-dim 
($x^{\pm},w$) part of the bulk see $z=\infty$ Lifshitz scaling.

It is interesting to ask if a constant-$x^+$ hypersurface has an
induced metric resembling that of a 4-dim $z=2$ Lifshitz spacetime.
Consider a static D5-brane probe stretched along $x^-,x_i,w$ and an
$S^2\in S^5$. This is effectively a domain wall in the $x^+$-direction
of the $AdS_5$ part of the bulk spacetime.
Since a constant-$x^+$ hypersurface is null, it is difficult to 
explicitly realize a $z=2$ $d=4$ Lifshitz spacetime as the induced 
metric on the 4-dim part of the D5-brane probe. Consider the bulk metric\
$ds^2 = {1\over w^2} [-2dx^+dx^- + w^2Q^2 (dx^+)^2 + dx_i^2]
+ {dw^2\over w^2} = \left(Qdx^+ - {dx^-\over Qw^2}\right)^2 
- {(dx^-)^2\over Q^2w^4} + {dx_i^2\over w^2} + {dw^2\over w^2}$ .
We see that on the slice\ $dx^+ = {dx^-\over Qw^2}$ ,
the induced metric is precisely $Lif^{z=2}_4$ times a compact 
space: however this is not a well-defined hypersurface.

The dual gauge theory is the 4-dim \Nf\ SYM theory living on flat 
spacetime, the boundary metric being flat, with the gauge coupling 
lightlike-deformed as\
$g_{YM}^2(x^+)=e^{\Phi(x^+)}\equiv g_s e^{2Qx^+}$, with $g_s=e^{\Phi_0}$. 
The linear-dilaton-like coupling gives a strong coupling Liouville-like
wall at one $x^+$-end, while for $x^+\ll 0$, the gauge theory becomes
weakly coupled and arbitrarily calculable perturbatively.
Correspondingly the spacetime ceases to be reliable in this regime,
where the string coupling becomes small. The string frame metric is\
$ds_{str}^2=e^{\Phi/2} ds^2$. This degenerates for $x^+\ra -\infty$,
where the curvatures become large.\\
The fact that the dilaton has a non-normalizable deformation turned on
means that the operator $Tr F^2$ is sourced. Since the deformation is
lightlike, there exist no nonzero contractions involving
$\del_+g_{YM}^2$ since there are no tensors with multiple upper
$+$-indices. Thus $Tr F^2$ continues to be a marginal dim-4 
operator\footnote{After this paper appeared on the arXiv, we were 
informed of \cite{Costa:2010cn} where some aspects of lightlike, 
or chiral, deformations of AdS/CFT have been discussed.}.

We will now calculate the 2-point correlation function in this 
linear dilaton case (\ref{metLinDil}):\ 
possible mode functions\ $\varphi(x)=e^{ik_-x^-+ik_ix^i} e^{g(x^+)} R(w)$ 
reduce the scalar wave equation\\ 
${1\over\sqrt{-g}} \del_\mu (g^{\mu\nu} \sqrt{-g} \del_\nu \varphi) 
- m^2\varphi = 0$\ \ to\ \
$-2ik_-g' + {w^3\over R(w)} \del_w \Big({1\over w^3} \del_w R(w)\Big) 
- k_i^2 - {m^2\over w^2} + w^2Q^2 k_-^2 = 0$.
With\ $g={-i\chi^2x^+\over 2k_-}$,\ the radial equation becomes
\be\label{LinDilRadEqn}
w^3\del_w \Big({1\over w^3} \del_w R(w)\Big) - 
\left(k_i^2 + \chi^2 + {m^2\over w^2} - w^2Q^2 k_-^2\right) R(w) = 0\ ,
\ee
Before we see this in detail, note that the boundary asymptotics of
this equation near $w=0$ show that the last term $w^2Q^2k_-^2$ in the
equation is subdominant and the mode functions approach those of $AdS$
in lightcone coordinates. Thus the 2-point function in this leading
approximation is\ ${1\over [(\Delta{\vec x})^2]^{\Delta}}$:\ in 
particular for points that are essentially on the same $x^+$-plane, 
\ie\ $\Delta x^+\ll \Delta x_i$, the 2-pt function exhibits spatial 
power law behaviour\ ${1\over [\sum_i (\Delta x^i)^2]^{\Delta}}$ .

In greater detail, the radial equation (\ref{LinDilRadEqn}) is 
exactly solvable in terms of confluent hypergeometric functions 
\cite{abramsteg}. Taking\ $R(w)=w^\Delta e^{\al w^2} f(w)$, and 
redefining\
\be\label{paramk2ala}
k^2=\chi^2+k_i^2\equiv -2k_+k_-+k_i^2 ,\quad\ \
\al=-{iQk_-\over 2}\ ,\quad\ \ \Delta=2+\sqrt{4+m^2} = 2+\nu , 
\ee
the radial equation becomes\ \
$z {d^2f\over dz^2} + (\Delta - 1 - z) {df\over dz} - 
({\Delta - 1\over 2} - {k^2\over 8\al}) f = 0$\ \ (where\ $z=-2\al w^2$)\ 
which is the confluent hypergeometric equation. This 
gives the momentum space bulk-to-boundary propagator\ 
(we have\ $\varphi(x^\mu,w)=\int d^4k\ \varphi(k) G(k,w)$) 
\bea
&& G(k_i,k_+,k_-, w) = {\cal N}(k)\ e^{ik_ix^i+ik_-x^-+ik_+x^+}\ 
w^\Delta e^{\al w^2} U(a,c,-2\al w^2)\ ,\nonumber\\ 
&& \qquad\ a = {\Delta - 1\over 2} - {k^2\over 8\al} 
= {\nu + 1\over 2} + {k^2\over 4iQk_-}\ ,\ ,
\qquad c=\Delta - 1 = \nu + 1 \ ,
\eea
where ${\cal N}(k)$ is a normalization factor which we will choose so 
as to set $G(k,\epsilon)=e^{ik_\mu x^\mu}$ on a cutoff surface $w=\epsilon$.
We have chosen the confluent hypergeometric function $U(a,c,z)$ in accord 
with the requirement of regularity in the interior ($w\ra\infty$) and 
the expectation that for $Q=0$, this should reduce to the standard
Bessel functions $w^2 K_\nu(kw)$ (radial part) for $AdS_5$ in
lightcone coordinates.  Now from the near-boundary ($w\sim 0$)
asymptotic form of the bulk-to-boundary propagator, we can identify
the momentum space 2-point correlation function as a ratio of the
growing and decaying (non-normalizable and normalizable) pieces.
For non-integral $\nu$ (\ie\ $c$), we have the confluent hypergeometric 
function asymptotics\ 
$U(a,c,z)\sim\ {\pi\over \sin \pi c} ({1\over\Gamma(1+a-c) \Gamma(c)} 
- {z^{1-c}\over\Gamma(a)\Gamma(2-c)} + \ldots)$. Then the momentum 
space 2-point correlation function can be read off as
\be\label{2ptfnNonintegralnu}
 \langle {\cal O}(k) {\cal O}(-k)\rangle = -\nu 2^\nu \al^\nu 
{\Gamma(-\nu)\over\Gamma(\nu)} {\Gamma(a)\over\Gamma(a-\nu)}\
=\ -\nu 2^\nu \al^\nu {\Gamma(-\nu)\over\Gamma(\nu)} 
{\Gamma({1+\nu\over 2} + {k^2\over 4iQk_-})\over
\Gamma({1-\nu\over 2} + {k^2\over 4iQk_-})}\ .
\ee

For $\nu$ integral, there are additional terms in the asymptotics 
of $U(a,c,z)$:\
the small-$w$ expansion of the radial part of the bulk-to-boundary 
propagator, after appending an overall normalizing factor so as to make 
$G(k_i,k_+,k_-,w=\epsilon)=e^{ik_\mu x^\mu}$, is
\bea
&& (-2\al)^2 \Gamma(a)\ w^4 e^{\al w^2}\ U(a,\nu+1,-2\al w^2)  \nonumber\\
&& \quad =\ (1 + \al w^2 + {\al^2 w^4\over 2} + \ldots) \cdot \\ 
&& \qquad\qquad \left( 
1 + w^4 {(-1)^{\nu+1} (-2\al)^\nu\over\nu(\Gamma(\nu))^2} 
{\Gamma(a)\over\Gamma(a-2)} \big(\log(-2\al w^2) + \psi(a)\big) 
+ \ldots \right)\ ,\nonumber
\eea
where $\psi(a)$ is the digamma function. 
After removing unimportant terms -- those removable by local 
counterterms and contact terms -- the momentum space 2-point correlation 
function (noting that $G$ vanishes in the interior $w\ra\infty$) 
can be read off from the boundary action\ 
$\int d^4k\ \varphi(k)\varphi(-k) (\sqrt{-g} g^{ww} G(-k,w) 
\del_w G(k,w))|_{w=\epsilon}$:\ 
it is basically the coefficient of the $w^4$-term in the second bracket 
(higher order terms vanish as $\epsilon\ra 0$).
In particular for the case of massless scalars with\ $\Delta=4, \nu=2$, as
\bea\label{2ptnu=2}
&&  \langle {\cal O}(k) {\cal O}(-k)\rangle\ \sim\ 
 4 (\al a - \al) (\al a - 2\al)\ (\log(2\al) + \psi(a)) \nonumber\\
&&\quad \sim\  \left((k_i^2-2k_+k_-)^2 + 4Q^2k_-^2 \right)\ 
\left(\log(iQk_-) + \psi\left({3\over 2} + {k_i^2-2k_+k_-\over 4iQk_-}
\right)\right). \qquad 
\eea
In the limit $Q\ra 0$, we have\ $\al\ra 0,\ a\ra\infty$, so that 
using the asymptotics\ $\psi(a)\sim \log a$ and (\ref{paramk2ala}), 
we recover the familiar $AdS_5$ correlator\ $k^4 \log k^2$ as a check. 
For other integral $\nu$ values, we obtain a correlator of the form\
$2^\nu (\al a - \al)\ldots (\al a - \nu\al) (\log(-2\al) + \psi(a))$
which asymptotes to $k^{2\nu}\log k^2$ in the $Q\ra 0$ (AdS) limit.

This calculation beginning with (\ref{LinDilRadEqn}) and the resulting
momentum space expressions (\ref{2ptfnNonintegralnu}) (\ref{2ptnu=2})
are structurally similar to those in \cite{Kachru:2008yh}.  In fact,
for $k_+\sim 0$, we see that the radial scalar equation
(\ref{LinDilRadEqn}), using (\ref{paramk2ala}), reduces precisely to
the equation for a scalar in the 4-dim $z=2$ Lifshitz background\
$ds^2=-{dt^2\over r^4} + {dx_i^2 + dr^2 \over r^2}$. Thus the
corresponding momentum space correlation functions
(\ref{2ptfnNonintegralnu}) (\ref{2ptnu=2}) (continued to Euclidean 
signature) are in fact identical to those discussed (in the Euclidean
calculation) in \cite{Kachru:2008yh}, as expected for the dimensional
reduction of the 5-dim AdS lightlike deformation to the 4-dim Lifshitz
one argued in \cite{Balasubramanian:2010uk,Donos:2010tu}\ (which we
expect to correspond to the $k_+\sim 0$ sector here).

For the case with the $x^+$-direction treated as noncompact, there are
more features here, due to the linear dilaton configuration which acts
like a potential in the $x^+$-direction.
(We note as an aside that this calculation also holds if we change 
$Q^2\ra -Q^2$, changing the sign of $g^{--}$: in this case 
$x^+$ becomes the natural time variable appropriate 
for this $z=0$ Schrodinger system and $k_-$ (coupling to $x^-$) 
is then a spatial momentum.) 
Some insight into the structure here is obtained by noting that\
\be
(k_i^2-2k_+k_-)^2 + 4Q^2k_-^2 = \left(k_i^2-2(k_+-iQ)k_-\right)\ 
\left(k_i^2-2(k_++iQ)k_-\right)\ .
\ee
Thus the effective $x^+$-momentum is shifted as\ $k_+\ra k_+\pm iQ$, 
and $e^{ik_+x^+}\ra e^{ik_+x^+} e^{\pm Qx^+}$. This is reminiscent of 
the Liouville-like wall in $c=1$ string theory 
\cite{Klebanov:1991qa,Ginsparg:1993is}.
Also note that for $k_i=0$, we have\ $(k_i^2-2k_+k_-)^2 + 4Q^2k_-^2
\ra (k_+^2+Q^2)k_-^2$, which gives an effective mass-gap in the 
$x^+$-direction. Another corroboration of this interpretation comes 
from looking at solutions to the wave equation in the SYM theory. 
The gauge field sector has the free action\ 
$S = \int d^4x {1\over g_{YM}^2(Qx^+)} Tr F^2$, which for the 
transverse components $A_i$ which are physical degrees of freedom 
in \eg\ lightcone gauge, becomes
$S_{A_i} = \int d^4x\ e^{-\Phi(x^+)} [ (\del_j A_i)^2 - 2(\del_+A_i)(\del_-A_i)]$, 
which is essentially two copies of a scalar moving in an 
$x^+$-dependent background.
Then the wave equation\ 
$e^{\Phi(x^+)}\del_+(e^{-\Phi(x^+)}\del_-A_i)+\del_-\del_+A_i+\del_j^2A_i=0$\
for modes of the form\ $e^{ik_+x^++ik_-x^-+ik_i x^i}$ gives\
$k_i^2 + 2(k_+ + iQ) k_- = 0$,\ \ie\ $k_+ = -{k_i^2\over 2k_-} - iQ$.\
The wave modes then become\ 
$e^{ik_ix^i+ik_-x^-+i{k_i^2\over 2k_-}x^+} e^{Qx^+}$, which are damped 
as $x^+\ra -\infty$.
Furthermore for generic $k_i,k_-$, we see that the $x^+$-momentum 
$k_+$ is nonzero, \ie\ generic waves will be forced to move 
along the $x^+$-direction due to the dilaton $x^+$-potential.
Admittedly this is in the free gauge theory while our calculation 
in the weakly coupled bulk geometry applies to a strongly coupled 
gauge theory: however the basic structure of the wave modes seems 
suggestive.

The scaling of the coefficient\ $(k_i^2-2k_+k_-)^2 + 4Q^2k_-^2$ in
(\ref{2ptnu=2}) is consistent with that of the $x_i,x^{\pm}$ in
(\ref{z=2infty}). To gain some insight into the $z=\infty$ scaling of
the $x^\pm$-directions, note that the dispersion relation $\omega\sim
k^z$ with $z\ra\infty$ can be rewritten as\ $k\sim \omega^{1/z}\sim 
\omega^0$. In the $x^\pm$-directions ($k_i=0$), we see from the
coefficient in (\ref{2ptnu=2}) that $k_+$ has a damping piece
independent of $k_-$, reflecting the $x^+$-potential given by the
dilaton.

The momentum space correlator can now be used to evaluate position 
space 2-point functions in certain limiting cases. For instance, as 
in \cite{Kachru:2008yh}, in the limit where the digamma function
$\psi(a)$ has dominant contribution and can be approximated by its 
leading constant term (for the argument of $\psi(a)$ being small), 
we can Fourier transform the coefficient\ 
$(k_i^2-2k_+k_-)^2 + 4Q^2k_-^2$\ using Schwinger parameters and find\ 
${e^{-Q\Delta x^+} \over [\Delta x_i^2 - 2\Delta x^+ \Delta x^-]^4}\ 
{1\over\Gamma(-1-\delta)}\int_0^1 dx\ {e^{-2Q\Delta x^+ x}\over x^2(1-x)^2}$,
after regulating it (the pole near $x=0$ cancels with corresponding 
terms in the $\Gamma$-prefactor). For $Q=0$, this is in fact just 
the correlator for $AdS_5$ in lightcone coordinates. For $Q\neq 0$, 
the last integral can be expressed in terms of incomplete 
$\Gamma$-functions.

Finally we note that we have calculated this 2-point correlation 
function by imposing certain boundary conditions on the solution 
to the bulk scalar wave equation, which are natural generalizations 
of those in $AdS_5$. Presumably one can find interesting 
real-time structure by studying various other boundary conditions 
on this system as discussed in \eg\ 
\cite{Hartnoll:2009sz,Herzog:2009xv,McGreevy:2009xe}.

The bulk arguments here for $x^+$-noncompact can also be made for
linear scalar deformations of $AdS_4\times X^7$ in M-theory, the
scalar arising from the flux components. The correlator then gives
some insight into the spatial structure of $2+1$-dim Lifshitz-like
field theories. It would be interesting to explore this further.

\section{On (non-geometric) DLCQ of a linear dilaton Lifshitz system}

We will now revert to compactifying the $x^+$-direction as in
\cite{Balasubramanian:2010uk}. The basic point there from the dual
gauge theory point of view is that a DLCQ of the \Nf\ SYM theory gives
a nonrelativistic theory with Galilean symmetries with dynamical
exponent $z=2$: then a varying coupling $g_{YM}^2=e^{\Phi(x^+)}$ along
the compact $x^+$-direction breaks the shift symmetry giving Lifshitz
symmetries. While a similar symmetries-level argument holds in the
bulk geometry too, a more detailed dimensional reduction appears
difficult, especially in terms of understanding if a Wilsonian
separation-of-scales argument holds. For instance, since $g_{++}\sim
(\Phi')^2$ contains $x^+$-dependence in general, the bulk 5-dim
geometry does not admit a standard Kaluza-Klein reduction.
We note however the exception\ $\Phi'=const$: this is a linear 
dilaton system that we have discussed in the previous section with 
the $x^+$-direction noncompact.

In this section, we would like to investigate whether we can
compactify the $x^+$-direction with such a linear dilaton
configuration, with a view to finding $AdS$ null deformations
(\ref{metLif5}) with $\Phi'=const$, admitting conventional
Kaluza-Klein reduction along the $x^+$-circle.

\begin{figure}
\bc
\epsfig{file=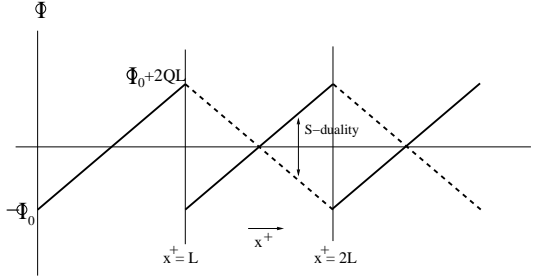, width=7cm}
\caption{{\small A linear dilaton configuration and compactification 
upto S-duality.}}
\label{figLinDil}
\ec
\end{figure}

With $\Phi'=const$, we have $\Phi=\Phi_0+2Qx^+$, \ie\ we have a linear
dilaton profile. A priori, this is in contradiction with the proposed
compactification of the $x^+$-direction: a dilaton linear for all
$x^+$ does not respect this. However let us consider a piecewise
linear dilaton configuration
\bea\label{linDilCompax}
\Phi &=& \Phi_0 + 2Qx^+ ,\qquad\qquad\qquad\  x^+ \in [0,L]\ ,\qquad\nonumber\\
&=& \Phi(L) - 2Q (x^+-L) ,\qquad x^+ \in [L,2L]\ ,\ \ \ldots 
\eea
This is a ``sawtooth''-like (piecewise) linear dilaton configuration 
$\Phi(x^+)$, plotted in Fig~\ref{figLinDil}.
$\Phi(x^+)$ is a continuous but not smooth function on the $x^+$-line: 
at the locations\ $x^+=L,\ 2L,\ \ldots$, we see that $\Phi'$ has a 
jump discontinuity from $+2Q$ to $-2Q$\ \ (to be precise, define\ 
$\Phi'=+2Q,\ \ 0<x^+\leq L$, and $\Phi'=-2Q,\ \ L<x^+\leq 2L$ , 
and so on). 
However we see that the Einstein metric is smooth since the dilaton 
$\Phi$ appears in the metric as\ $g_{++}\sim (\Phi')^2$. Thus all 
metric properties (curvature, geodesics and so on) are smooth also.
The bulk Einstein metric is
\be\label{metLinDilnonGeom}
ds^2 = {1\over w^2} [-2dx^+dx^- + dx_i^2 + w^2 Q^2 (dx^+)^2 + dw^2]\ .
\ee

It is desirable to demand that $\Phi(x^+)$ be a continuous function on
the unwrapped $x^+$-circle, \ie\ on the $x^+$-line: this is true for
the dilaton configuration (\ref{linDilCompax}) above. However the
dilaton $\Phi(x^+)$ is not periodic and does not therefore respect the
compactification along the $x^+$-direction.  We would like to
understand if this dilaton configuration can be somehow made to
respect the compactification, \ie\ $\Phi(x^++kL)=\Phi(x^+)$, for any
$k\in \BZ$.\\
This does not appear to be possible in conventional supergravity per
se. However note that we have at our disposal the exact S-duality
symmetry of the IIB string theory. It is therefore interesting to ask if we
can use this to construct a solution with the dilaton $\Phi(x^+)$
periodic on the $x^+$-circle, possibly along the lines of
non-geometric string constructions.  Supergravity solutions involving
nongeometric constructions with nontrivial winding around U-duality
orbits have been studied previously in \eg\
\cite{Greene:1989ya,Vafa:1996xn,Kumar:1996zx,Hellerman:2002ax,
Dabholkar:2002sy}. The S-duality in question here makes
our construction nonperturbative, in contrast to some of these
constructions that involve T-duality. However, the backgrounds in
question here are considerably simpler in some ways: the absence of a
nontrivial axion means that there are no nontrivial 7-brane sources of
the sort that arise in F-theory constructions. Nor are there 
singularities from degenerations of the fibre.

S-duality is the symmetry $\tau \ra -{1\over\tau}$, with
$\tau=c_0+ie^{-\Phi}$ the complexified axion-dilaton coupling. Since
the axion is trivial in this background, the S-duality symmetry
reduces to strong $\ra$ weak coupling duality,\ \ie\ $\Phi\ra -\Phi$.\\
Using this, we see that it suffices to require \eg\ 
$\Phi(x^++L)=-\Phi(x^+)$. From (\ref{linDilCompax}), at the end of 
the interval $x^+\in [0,L]$, we have\ $\Phi(L)=\Phi_0+2QL$, which 
cannot equal $\Phi(0)=\Phi_0$. However since $\Phi(0)\equiv -\Phi(0)$ 
upto S-duality, consider requiring $\Phi(L)=-\Phi(0)$: this is 
consistent since
\be
\Phi(L)=-\Phi(0)\quad \Rightarrow \qquad 
2\Phi_0=-2QL ,\quad i.e.\ \quad g_s=e^{\Phi_0}=e^{-QL}\ .
\ee
This gives\ $\Phi(x^+)=2Q(x^+-{L\over 2})$ for $x^+\in [0,L]$. We note 
that $\Phi=0$ at $x^+={L\over 2}$, with $\Phi(x^+)$ being antisymmetric 
about this midpoint.\\
Now consider the interval\ $x^+\in [L, 2L]$: from (\ref{linDilCompax}), 
we have\ $\Phi(x^+)=\Phi(L)-2Q(x^+-L)=-\Phi_0-2Q(x^+-L)$. We see that on 
this interval,\ $\Phi(x^+)\equiv -\Phi(x^+)=\Phi_0 + 2Q(x^+-L)$. Thus 
the S-duality symmetry $\Phi\ra -\Phi$ amounts to flipping a local 
linear piece about the $x^+$-axis, \ie\ flipping a downward sloping 
linear piece to an upward sloping one. Thus comparing the dilaton 
values at $x^+=a\in [0,L]$ and $x^+=L+a \in [L, 2L]$ using 
(\ref{linDilCompax}), we see 
\be
\Phi(L)=-\Phi_0 \quad \Rightarrow\quad 
(\Phi_0+2Qx^+)|_a = -(\Phi(L) -2Q(x^+-L))|_{L+a}\ .
\ee
In other words, upto S-duality, the  piecewise linear dilaton 
configuration configuration is $x^+$-periodic if the linear pieces 
are antisymmetric about the $x^+$-axis, \ie\ if 
$\Phi_0=\log g_s = -QL$. This is easy to see pictorially 
(Fig~\ref{figLinDil}).

Since the dilaton is periodic upto S-duality in this manner, the 
string frame metric appears to be well-defined. It is unclear to 
us at this point if there is a geometric way to interpret this 
resulting nongeometric construction along the lines of \eg\
\cite{Vafa:1996xn,Kumar:1996zx}.

Since the asymptotic value of the string coupling is not arbitrary but
fixed by this nongeometric construction as $g_s=e^{\Phi_0}=e^{-QL}$,
one could potentially worry if our solution is reliable and whether
stringy corrections to this geometry are becoming important.  In this
regard, note firstly that the dilaton is always bounded so that there
are no apparent singularities that wreck the solution anywhere (in
contrast the points on moduli space where the elliptic fiber
degenerates correspond to the locations of the 7-brane singularities
in F-theory). Furthermore note that the solution preserves half
supersymmetry, suggesting the absence of various corrections. The
supersymmetry of these solutions is closely related to the lightlike
nature of these solutions, which suggests that higher derivative
corrections to the solution in fact vanish, \ie\ that these are 
exact string backgrounds (perhaps similar to $AdS_5\times S^5$\
\cite{Kallosh:1998qs}). We have seen that a way to uplift the
Lifshitz-like symmetries of the 4-dim Lifshitz system to the 5-dim
system is to turn on lightlike deformations of the form
(\ref{metLinDilnonGeom}) which reflect these Lifshitz-like symmetries.
Thus the requirement of Lifshitz-like symmetries is effectively
captured by the lightlike nature of these solutions, which would seem
to preclude corrections to the higher dimensional solution. These
arguments suggest that this nongeometric solution is reliable, insofar
as standard dimensional reduction is concerned: the low energy or
Einstein metric is now $x^+$-independent and admits a conventional
Kaluza-Klein reduction, giving rise to the $z=2$ Lifshitz 4-dim
spacetime. It would be interesting to investigate these arguments
further with a view to understanding how robust they are.

The gauge theory dual in this case has the gauge coupling varying as\
$g_{YM}^2=g_s e^{2Qx^+}$:\ the coupling is always bounded on the
$x^+$-circle. The gauge theory also has an exact S-duality symmetry 
and our bulk construction effectively implies a non-geometric 
construction of the gauge theory too. The gauge coupling then is 
periodic upto S-duality, as we have argued for the dilaton. It would 
be interesting to explore this further.

\subsection{Solutions with lightlike axion}

It is worth mentioning that there are very similar solutions sourced
purely by a lightlike Type IIB axion $c_0=c_0(x^+)$ too: these are of 
the form
\be
ds^2 = {1\over w^2}[-2dx^+dx^- + dx_i^2 + w^2 (\del_+c_0)^2 (dx^+)^2 + dw^2]
+ ds_5^2\ , \qquad c_0=c_0(x^+)\ .
\ee
Specializing to a linear axion configuration, we have
\be
ds^2 = {1\over w^2}[-2dx^+dx^- + dx_i^2 + w^2 Q^2 (dx^+)^2 + dw^2] + ds_5^2\ ,
\qquad c_0=c_0^0+2Qx^+\ ,
\ee
which is akin to (\ref{metLinDil}) except with a constant dilaton $\Phi$. 
The existence of these solutions can be directly seen from the IIB
supergravity equations of motion or alternatively by restricting to a
nontrivial axion alone in the solutions in \cite{Donos:2010tu}. The
axion equation of motion is automatically satisfied due to the 
lightlike nature.

With the $x^+$-direction treated as compact, we expect then that
similar nongeometric constructions can be performed on these linear
axionic solutions too, with perhaps more similarity to F-theory
constructions.  In this case, $c_0(x^+)=c_0^0+2Qx^+$ would shift as
$c_0\ra c_0+2QL$ under $x^+\ra x^++L$. This is then equivalent to the
$\tau\ra\tau+1$\ shift if\ $QL={1\over 2}$. The dilaton in this case is
constant.  Presumably there exist solutions of this sort with
both axion-dilaton nontrivial.

These axionic solutions are reminiscent of the D3-D7 solutions in
\cite{Azeyanagi:2009pr}. The axion gives rise to a $\theta$-angle term
in the dual gauge theory. It would be interesting to explore the
interpretation of these solutions further.

\section{Discussion}

We have discussed certain lightlike deformations of $AdS_5\times X^5$
sourced by a lightlike dilaton $\Phi(x^+)$ dual to the \Nf\ SYM theory
with a lightlike varying gauge coupling, building on
\cite{Balasubramanian:2010uk}, and argued that in the case with $x^+$
noncompact, these solutions describe anisotropic Lifshitz-like systems
with $z=2$ and $z=\infty$ scaling in the $x^-,x_i$- and
$x^{\pm}$-directions respectively, alongwith a spatial $x^+$-potential
stemming from the dilaton. We have then focussed on linear dilatonic
systems and studied 2-point correlation functions of operators dual to
bulk scalar modes in these cases. We then discussed a certain
nongeometric string construction to compactify the $x^+$-direction. We
have also pointed out similar axionic solutions.

Our bulk discussion with $x^+$ noncompact readily generalizes to
$AdS_4\times X^7$ solutions in 11-dim supergravity, building on the
corresponding $z=2$ solutions in \cite{Balasubramanian:2010uk}. The
dual field theory is not entirely clear to us: it would be interesting
to explore this further with a view to identifying possible lightlike
deformations of Chern-Simons theories arising on M2-brane stacks.

The linear dilaton system we have discussed is reminiscent of $c=1$
string theory and also NS5-branes where a linear dilaton arises:
perhaps there are interesting nonrelativistic systems involving these.

The linear dilaton example illustrates the spatial structure of these
sorts of theories as we have seen, with the dilaton acting as a
potential in a sense. We expect that more general solutions
(\ref{metLif55}) will exhibit similar features too, although studying
observables such as correlation functions might be more intricate.
However it is interesting to note some aspects of certain specific
solutions: since the spacetimes (\ref{metLif55}) are essentially a
family of solutions for any $\Phi$, distinct solutions exhibit various
interesting features.  An interesting solution is obtained by taking
$\Phi'=2Q\tanh (Qx^+)$, giving
\be
ds^2 = {1\over w^2}[-2dx^+dx^- + dx_i^2 + w^2 Q^2 \tanh^2(Qx^+) (dx^+)^2 
+ dw^2]\ , \qquad  \Phi=\Phi_0+2\log\cosh(Qx^+)\ .
\ee
Note that $\Phi'\ra \pm Q$ as $x^+\ra \pm\infty$.  Also as $x^+\ra 0$, 
we have $\Phi'\ra 0$. Thus the bulk spacetime is Lifshitz-like away 
from $x^+=0$, a ${1\over Q}$-sized region near which the spacetime is
approximately $AdS$-Schrodinger with $g_{++}\ra 0$. This suggests 
that this solution constitutes a ``junction'' of two $z=2$ 
Lifshitz-like systems joined together with an $AdS$-Schrodinger-like 
core about $x^+=0$. For $Q$ large, the size of this core shrinks and 
the ``junction'' becomes sharper.
The gauge coupling for the dual \Nf\ SYM theory in this case is\ 
$g_{YM}^2=e^\Phi=g_s \cosh^2(Qx^+)$. The fact that the string (or 
gauge) coupling runs along $x^+$ implies a varying D-probe tension 
along that direction, \ie\ there is a potential for D-brane probes 
(\ie\ charged dyonic states in the gauge theory) along the 
$x^+$-direction, with a maximum for a D-brane probe tension 
${1\over g_s}$ at $x^+=0$.

Another interesting system arises for\ $\Phi=\Phi_0+\tanh Qx^+$. Then 
the metric becomes
\be
ds^2 = {1\over w^2}[-2dx^+dx^- + dx_i^2 + w^2 {Q^2(dx^+)^2\over\cosh^4(Qx^+)}
+ dw^2]\ , \qquad  \Phi=\Phi_0+\tanh(Qx^+)\ .
\ee
Now for large $x^+$, we see that $g_{++}\ra 0$ and we have a 
Schrodinger system there, while in the vicinity of $x^+=0$, this 
resembles the linear dilatonic system discussed previously, with 
Lifshitz-like behaviour. The dual field theory reflects this, with 
an interpolation between Galilean and Lifshitz-like regimes.

In a sense, these spacetimes exhibit holographic interpolations in the
$x^+$-direction between asymptotic $z=2$ Lifshitz-like and
Schrodinger-like regions. It would be interesting to find solutions
generalizing these where there is additional radial dependence,
perhaps along the lines of \cite{Nishioka:2010ha}, and obtain a more
detailed understanding of holographic renormalization group flows
between these non-relativistic systems.

It is tempting to speculate that solutions with $x^+$-noncompact where
the coupling $g_{YM}^2=e^{\Phi(x^+)}$ is periodic in $x^+$ would
appear effectively lattice-like with a periodic potential for charged
dyon states. Likewise solutions with $e^{\Phi(x^+)}$ possessing some
randomness might simulate disorder along the $x^+$-direction. It might
be interesting to explore these further.


\vspace{5mm}
\noindent {\small {\bf Acknowledgments:} It is a great pleasure to 
thank K. Balasubramanian, A. Dabholkar, S. Das, S. Kalyani, 
S. Minwalla, M. Rangamani, B. Sathiapalan, A. Sen, N. Suryanarayana 
and especially S. Trivedi for helpful conversations. I also thank 
K. Balasubramanian for collaboration in the early stages of 
this work. I thank the organizers of the ISM 2011 string theory 
workshop, Puri, for hospitality while this work was in progress. 
This work is partially supported by a Ramanujan Fellowship, DST, 
Govt of India.}

\end{document}